\documentclass[manuscript]{aastex}
\begin{document}

\title{Outflowing Diffuse Gas in \\ the Active Galactic Nucleus of NGC~1068}

\author{T. R. Geballe and R. E. Mason \altaffilmark{1}}
\author{T. Oka\altaffilmark{2}}

\altaffiltext{1}{Gemini Observatory, 670 N. A`ohoku Place, Hilo, HI, 96720, USA} 
\altaffiltext{2}{Department of Astronomy and Astrophysics and  Department of Chemistry, The Enrico Fermi Institute, University of Chicago, 5735 South Ellis Avenue, Chicago, IL 60637, USA}

\begin{abstract}

Spectra of the archetypal Type II Seyfert galaxy NGC 1068 in a narrow wavelength interval  near 3.7~$\mu$m have revealed a weak absorption feature due to two lines of the molecular ion H$_3^+$.  The observed wavelength of the feature corresponds to velocity of -70 km~s$^{-1}$ relative to the systemic velocity of the galaxy, implying an outward flow from the nucleus along the line of sight. The absorption by 
H$_3^+$ along with the previously known broad hydrocarbon absorption at 3.4~$\mu$m probably are formed in diffuse gas that is in close proximity to the continuum source, i.e. within a few tens of parsecs of the  central engine. Based on that conclusion and the measured
H$_3^+$ absorption velocity and with the assumption of a spherically symmetric wind we estimate a rate of mass outflow from the AGN of $\sim$1~M$_{\odot}$~yr$^{-1}$. 

\end{abstract}


\section{Introduction}

Active galactic nuclei (AGNs) contain a number of components on various spatial scales: the accretion disk, broad-line region (BLR), torus, narrow-line region (NLR), etc. Rather than being the static, well-defined structures that these terms may imply, in reality they are probably complex, dynamic, and in close interaction with each other and with the AGN's host galaxy. For example, spectroscopy of the NLR in some high-redshift quasars suggests that an AGN-driven outflow can suppress star formation in the host galaxy \citep{can12}. On much smaller scales, the dusty torus of the AGN unified model may be produced by gas flowing towards the AGN and play a central role in regulating accretion \citep{hop12}. Conversely, others have suggested that the torus originates in gas flowing out from the AGN (Elitzur \& Schlosman 2006), or is formed by the interplay between inflow and outflow (Wada 2012). Understanding the motions of gas in AGNs may therefore reveal clues about the way in which AGN and their surroundings interact with each other and evolve over time.

One might expect that where there is a dusty torus there are molecules and that therefore, for a heavily obscured Type II AGN, infrared spectroscopy of the central source would reveal absorption lines from molecules in or associated with the obscuring toroid, thereby probing the kinematics of the material on the line of sight to and close to the central engine. The most extensive  such spectroscopy to date has been of the archetypal Seyfert II galaxy, NGC~1068. However, those observations, at a variety of spectral resolutions \citep{lut04,mas06,geb09}, failed to detect lines of what should be the strongest infrared molecular band, the fundamental vibration-rotation band of carbon monoxide (CO) near 5~$\mu$m. \citet{geb09} concluded that the gaseous environment directly in front of NGC 1068's AGN, as viewed at mid-infrared wavelengths, is more similar to diffuse clouds, which have densities of 10-300 cm$^{-3}$, than dense clouds, which have higher densities. This conclusion was based not only on the absence of CO, but also on the presence of the 3.4-$\mu$m hydrocarbon absorption feature in NGC~1068 \citep{bri94}, which is not found in Galactic dense clouds and is  a reliable signature of a diffuse cloud environment. In Galactic diffuse clouds typically half of the hydrogen is in molecular form (H$_2$), but only about one percent of the carbon is in CO. A similar percentage of carbon in CO on the sightline to NGC~1068 would be consistent with its non-detection there \citep{geb09}. 

Unlike emission spectroscopy, absorption spectroscopy samples material only directly on the line of sight to a background continuum source.  At 3--5~$\mu$m the bright central source of NGC~1068 has an unresolved or marginally resolved core of linear dimension less than 0\farcs05 (3.5~pc at the 14.4 Mpc distance to NGC~1068) and a complex surrounding morphology of characteristic dimension $\sim$1\arcsec\  \citep{gra06}. When observed at 3-5~$\mu$m in good seeing on large telescopes without the benefit of adaptive optics the nuclear source has an effective full width at half maximum (FWHM) of $\sim$ 0\farcs3 (20~pc) in the N-S direction \citep{geb09}. The galactic disk of NGC~1068 is inclined at 40$^{\circ}$ such that the near side of the minor axis of the galaxy lies to southeast \citep[][see also de Vaucouleurs 1958]{dev91}; thus the sightline to the AGN passes through only a small portion of the galaxy within the nuclear region, and does not pass through the disk of the galaxy.

Combined knowledge of the detailed morphology and kinematics of the gas in the vicinity of an AGN is important for understanding the infall and/or outflow close to the central engine.  For NGC~1068, \citet{gar14} have used the Atacama Large Millimeter Array (ALMA) to observe the molecular gas and dust in this region, finding that the bulk of each is located in a circumnuclear disk (CND) of outer radius $\sim$200~pc whose center is offset by a few tens of parsecs from the position of the AGN. From the profiles and intensity ratios of some of the submillimeter lines in the CND they have identified the kinematic signature of a massive outflow, with a rate of many tens of solar masses per year. They identify the gas responsible for this signature as disk material being entrained by the ionized wind from the AGN. Closer to the AGN, on scales of 3--30 pc from the center, \citet{mul09} have found that the kinematics and morphology of the 2.12 ~$\mu$m H$_2$ line emission indicate linear streamers of gas are fueling the nucleus, and thus presumably the outflow. They interpret some of this gas as being situated on the near side of the nucleus. 
A contrasting interpretation of the H$_2$ line emission is given by \citet{bar14}, based on their velocity channel maps in the inner few hundred parsecs, obtained at a resolution of 8~pc. They find that their H$_2$ data are consistent with a radially expanding ring located in the plane of the galaxy. Regardless of the interpretation it is likely that the H$_2$ line emission arises in a small fraction of the molecular gas that is at high temperature immediately following shock-excitation or perhaps X-ray heating,  and whose motions are not necessarily representative of the kinematics of the neutral nuclear gas as a whole.

The CND observed by \citet{gar14} overlaps the AGN on the sky, so it is important to know whether it passes in front of, behind, or through the AGN. The nucleus of NGC~1068 is distinguished by a prominent bicone of ionized gas at a position angle of $\sim$10$^{\circ}$, extending several arc seconds on either side of the center. The ionization cone extending NNE from the AGN is much more prominent at optical wavelengths than the southern cone. Both that and the blue-shifts of the emission lines observed in it indicates that the NNE cone is tilted toward the sun. The plane of the CND is expected to be roughly perpendicular to the bicone. If so then the CND lies largely or entirely behind the AGN where they overlap on the sky. Nevertheless, in view of the huge quantity of molecular gas and dust that are present in the CND as well as the asymmetric location of the CND with respect to the central engine it remains surprising that until now no molecular absorption lines had been seen on the line of sight to the AGN.   

Clearly, an observation of absorbing gas on the line of sight to the AGN could provide important additional information about gas motions close to the central engine. Because the search for absorption lines of CO has been unsuccessful, an alternative infrared probe of the foreground gas column is needed. One possible candidate species is H$_3^+$. Although it is obviously an ion, H$_3^+$ is created by the reaction H$_2$~+~H$_2^+$ $\rightarrow$ ~H$_3^+$ ~+~H and resides in largely neutral and at least partially molecular gas. Vibration-rotation lines of H$_3^+$, although generally weak, are readily detectable not only in dense clouds \citep{geb96}, but also in numerous Galactic diffuse clouds \citep{mcc02, ind07}, including e.g. Cygnus OB2 No. 12 \citep{geb99}, whose 3.4-$\mu$m feature is less than half the strength of that toward NGC~1068 \citep{whi97,geb09}. Unless the lines of H$_3^+$  are greatly broadened toward NGC~1068 one would expect to detect them on the line of sight to the AGN. The close pair of vibration-rotation lines (1-0 $R$(1,1)$^u$ and 1-0 $R$(1,0)), at rest wavelengths of 3.66808~$\mu$m and 3.66852~$\mu$m, respectively, which arise from the lowest levels of para and otho H$_3^+$, are the best candidates for detection. In both diffuse and dense clouds in the Galactic plane these are the only two levels that are appreciably populated and the doublet is the strongest H$_3^+$ absorption feature.

\section{Observations and Data Reduction}

$L$-band spectra of NGC 1068 were obtained at the Frederick C. Gillett Gemini North Telescope on UT 2011 July 24 and October 11 and on UT 2014 November 28 and December 3, for programs GN-2011B-Q-53 and GN-2014B-Q-57, respectively.  Both sets of observations used the Gemini North Infrared Spectrograph (GNIRS), but in different configurations. The 2011 spectra were obtained using the 111 l/mm grating, short focal length camera and the 0.45 \arcsec\ wide slit and covered 3.49-3.79~$\mu$m. The 2014 spectra were obtained through the same slit, but used the long focal length camera and the 111 l/mm grating, which covered 3.64-3.74~$\mu$m and were heavily oversampled in wavelength. In each case the resolution at the wavelength of the sought after H$_3^+$ lines was 0.00082~$\mu$m (70~km~s$^{-1}$). The observations were obtained in the standard stare/nod-along-slit mode, with a nod of +/- 3\arcsec. The slit was oriented along the axis of the bicone at position angle 15$^{\circ}$ (E of N). The total exposure times were 64 minutes in 2011 and 120 minutes in 2014. Early type telluric standard stars were observed at closely similar air masses to NGC~1068 either before or after each set of observations. The measurements were made in photometric or near photometric conditions and in good seeing.

The initial stage of data reduction included flat-fielding, spike removal, extraction of spectra over 0\farcs75  long regions of the slit, wavelength calibration (using telluric lines and accurate to 0.0001~$\mu$m, corresponding to 10~km~s$^{-1}$ ), and ratioing by the spectra of the telluric standards. Normalized ratioed spectra were then shifted to compensate for differences in the earth's orbital motion relative to the coordinates of NGC~1068 on the different observing dates, and then combined to produce one spectrum for each of the two programs. The continuum of each spectrum had slight curvature, which we suspect is an instrumental effect due to differences in the distribution of light from NGC~1068 and the standard stars  and different positioning of them in the slit; to remove it and obtain flat spectra each spectrum was  divided by a spline fit to its continuum. The resultant spectra were then resampled on a common wavelength scale and coadded with equal weighting (as their noise levels are similar) in the narrow wavelength interval encompassing the expected wavelength of the H$_3^+$ doublet, near 3.682~$\mu$m. Strong telluric absorption lines are present on either side of this wavelength, but the interval 3.675-3.686~$\mu$m, where the redshifted H$_3^+$  doublet is expected to fall, is devoid of significant interference. The spectra from semesters 2011B and 2014B and the final combined spectrum are shown in Fig. 1. 

\begin{figure}[h!]
\begin{center}
\includegraphics[width=1\textwidth]{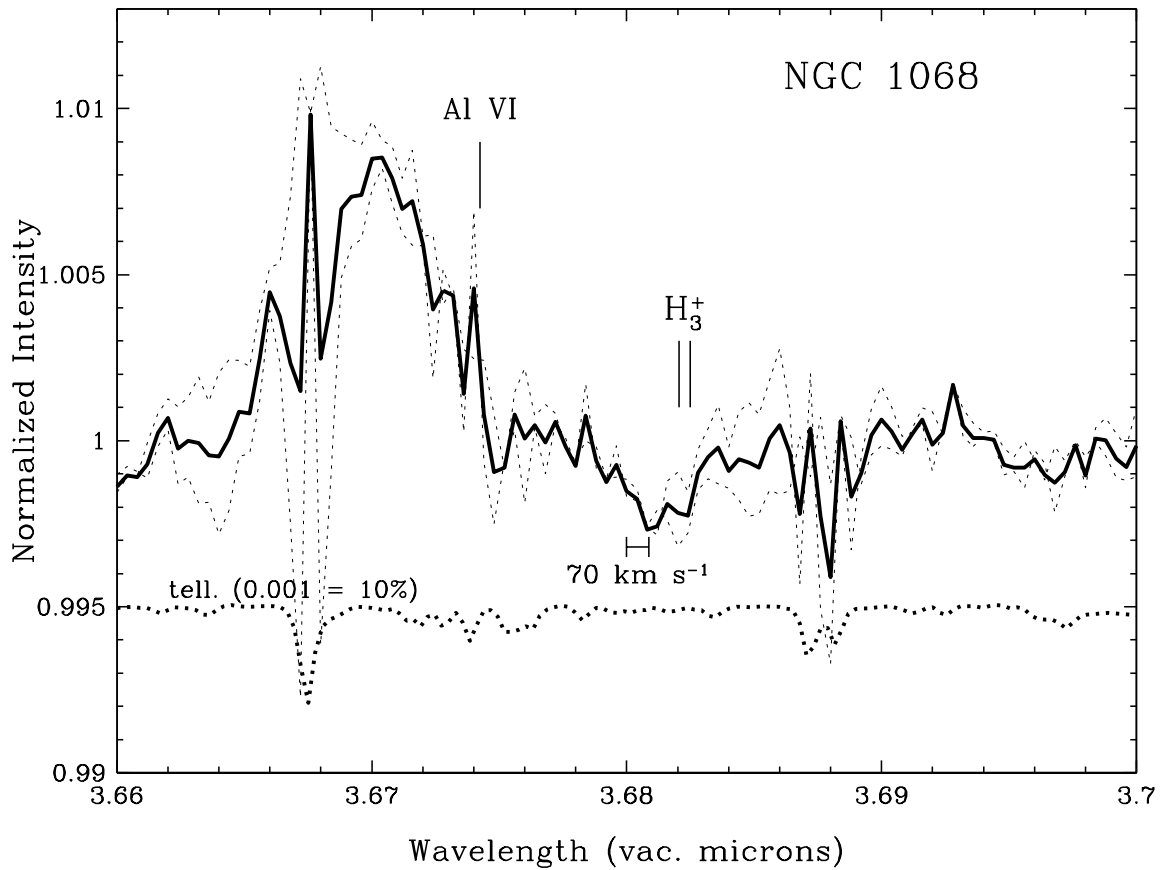}
\caption{Spectrum of the infrared nuclear source in NGC~1068 near the H$_3^+$ para and ortho line pair, at a resolution of 70 km s$^{-1}$.  The spectra from 2011 and 2014 (light dotted lines) and the combined spectrum (dark line) are shown, as is a 100 times scaled down spectrum of the atmospheric transmission (dark dotted line).The wavelengths of the H$_3^+$  lines and of the adjacent [Al\,{\sc vi}] coronal line are indicated  for a systemic heliocentric velocity of 1142 km s$^{-1}$ (see text). }
\label{spec}
\end{center}
\end{figure}

\section{Results}

The combined spectrum in Fig. 1 shows a detection of a very weak absorption feature with its centroid at 3.6815~$\mu$m, which we identify as the  blended pair of lines of para and ortho H$_3^+$ mentioned above.  A coronal emission line of [Al\,{\sc vi}], previously reported by \citet{geb09} is also present in the spectrum. Based on the fluctuations in the adjacent continuum in the combined spectrum the detection of the H$_3^+$ feature is at the 6$\sigma$ confidence level. However, the differences in the 2011 and 2014 spectra on the long wavelength portion of the continuum suggest that the true confidence level is not quite that high. In estimating the continuum and the noise (the latter based on point-to-point fluctuations in the spectrum) we ignored the spectral intervals near the telluric lines at 3.674~$\mu$m and 3.688~$\mu$m, as the noise levels there (which are dominated by fluctuations in the sky background) are significantly higher than elsewhere. 

Assuming a systemic heliocentric velocity of 1142~km~s$^{-1}$ (from the H~I Parkes All Sky Survey, as reported in the NASA/IPAC Extragalactic Database), the centroid of the absorption is blue-shifted by approximately 70 km~s$^{-1}$ from the systemic velocity of NGC~1068, indicating that the gas associated with the H$_3^+$ is flowing outward from the nucleus.   The feature has a measured equivalent width of  5.2~$\pm$~0.8~$\times~$~10$^{-6}$~$\mu$m and an observed FWHM of $\sim$150 km~s$^{-1}$.  Assuming that the two lines, which are separated by 36 km~s$^{-1}$, contribute equally to the feature, as they approximately do in Galactic diffuse clouds, the combined column density in the (1,1) and (1,0) levels is 7~$\times$~10$^{14}$~cm$^{-2}$ and the intrinsic FWHM of a single absorption line is approximately 100 km~s$^{-1}$. That width is considerably narrower than the more highly blue-shifted  [Al\,{\sc vi}]  line in Fig.~1. \citet{geb09} found that the peak intensity of the infrared coronal lines occurs a few tenths of an arc-second to the north of the infrared continuum peak. The present data also show that separation. Clearly the gas responsible for the absorption by H$_3^+$ and the gas in which the line emission by [Al\,{\sc vi}] arises are in completely different locations and physical environments. 

Other key lines of H$_3^+$ that could confirm the identification of the 3.6815-$\mu$m feature as due to H$_3^+$ and could provide constraints on the temperature and density of the material in which it originates are the $R$(1,1)$^l$ and $R$(3,3)$^{l}$ lines at 3.71548~$\mu$m and 3.53366~$\mu$m, respectively. Unfortunately each of these lines is likely to be weaker than the unresolved doublet and each is shifted into a wavelength interval where strong telluric absorption lines are present.  Despite the lack of confirmatory evidence we are confident in the identification of H$_3^+$, as (1) there are no other likely identifications for the feature and (2) the line pair was expected to be readily detected. For example, if the equivalent width of the doublet in NGC~1068 scaled as the ratio of equivalent widths of the observed 3.4~$\mu$m and doublet in the Galactic center (see Fig. 2), the equivalent width of the doublet in NGC~1068 would be 3~$\times$~10$^{-5}$~$\mu$m, six times greater than that observed. The surprising weakness of the doublet in NGC~1068 is discussed below. 

\section{Discussion}

\subsection{Extinctions to and Locations of Continuum and Absorption Feature}

$L$-band spectra of NGC~1068 have yielded two absorption features, a prominent 3.4~$\mu$m feature generally ascribed to aliphatic hydrocarbons, and a weak absorption due to a pair of closely spaced lines of H$_3^+$. Understanding where on the line of sight these absorptions occur requires first of all knowledge of the location of the source or sources of the infrared continuum. The adaptive optics images of \citet{gra06} show that within the portions of the slit from which the spectra of these features were extracted, by \citet{geb09} for 3.4~$\mu$m feature (0\farcs2~$\times$~1\farcs0) and by us for the H$_3^+$ feature (0\farcs45~$\times$~0\farcs75), the continuum is produced by a high surface brightness core and a complex morphology of more extended and lower surface brightness regions with the two being roughly comparable in total signal. 

Whatever the explanation for the $L$-band morphology; the likelihood is that the continuum is thermal emission from clouds containing hot dust \citep[e.g.,][]{gra06}, presumably heated by the central engine. If so, the continuum probably arises in the vicinity of each $\tau$=1 (in the $L$ band) surface in this complex region; in other words at distances of a few parsecs to a few tens of parsecs from the engine.  One can also estimate the mean optical depth to the $L$-band continuum from the depth of the 3.4~$\mu$m feature, assuming the relationship between the depth of the feature and the optical extinction in Galactic diffuse clouds \citep{god12} holds for NGC~1068, where the optical depth of the feature is 0.08 \citep{geb09}. This assumption yields a visual extinction of $\sim$21~mag, which, for normal extinction curves  (e.g., extinction proportional to $\lambda^{-1.7}$) implies an $L$-band optical depth close to unity. The near equality of the optical depths derived in these different ways is either a coincidence or, as we suspect, an indication that the dust producing the $L$-band continuum and the carrier of the 3.4~$\mu$m absorption (and by association the absorbing H$_3^+$) are physically close to one another, placing the absorbing material within a few tens of parsecs of the central engine. This proximity is not surprising in view of the orientation of the disk of the galaxy, as discussed earlier.

 \begin{figure}[h!]
\begin{center}
\includegraphics[width=1\textwidth]{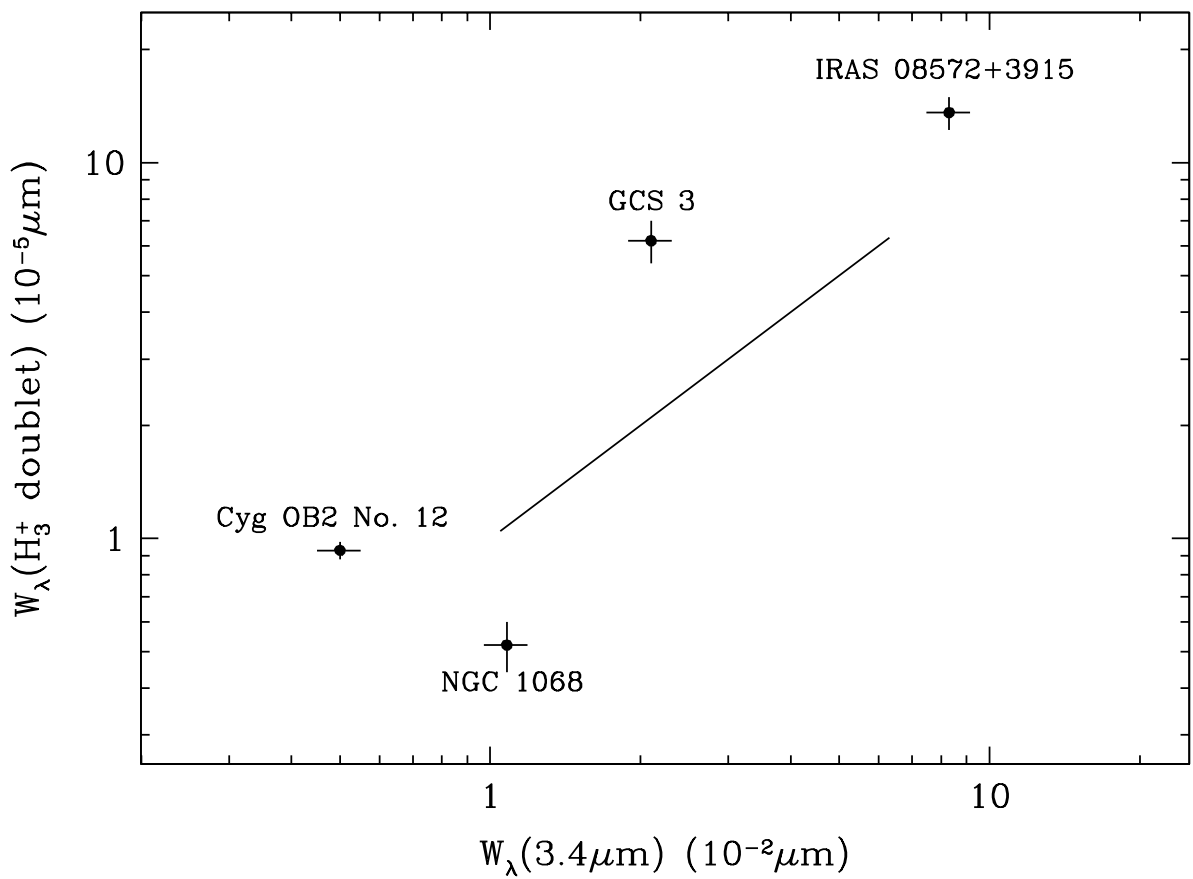}
\caption{Equivalent width of that portion of the absorption by the H$_3^+$ para-ortho line pair ascribed to diffuse gas vs. equivalent width of the 3.4~$\mu$m feature, for Cygnus OB2 No. 12, the Galactic center Quintuplet source(s) GCS~3, the ultraluminous infrared galaxy (ULIRG) IRAS 08572+3915, and NGC~1068.  Error bars are $\pm$1$\sigma$. The diagonal line is a one-to-one relationship between the two equivalent widths. The equivalent width of the H$_3^+$ feature in GCS~3 is based on the measurement toward GCS~3-2. In all but GCS3-2 the absorption by H$_3^+$ is assumed to be formed entirely in diffuse gas. In GCS~3-2 the contribution by H$_3^+$ in dense clouds has been subtracted. See text for details.}
\label{wlambda}
\end{center}
\end{figure}

\subsection{Low H$_3^+$ Equivalent Width}

Figure 2 is a plot of the equivalent width of that portion of the H$_3^+$ line pair arising in diffuse gas versus the equivalent width of the 3.4-$\mu$m absorption feature, for four sources: the Galactic plane source Cyg OB2 No. 12, the Galactic center's tight cluster of four sources known as GCS~3 (of which the brightest component is GCS~3-2), the ultraluminous infrared galaxy (ULIRG) IRAS 08572+3915, and NGC~1068.  The  H$_3^+$  equivalent widths for the first three sources are obtained from data or figures in \citet{geb99}, \citet{oka05}, and \citet{geb06}, respectively. The equivalent widths of the 3.4-$\mu$m feature were measured from spectra in \citet{san91}, \citet{chi00}, \citet{geb06}, and \citet{mas06}, respectively. For Cyg~OB2 No. 12, IRAS 08572+3915, and NGC~1068 either all of the  H$_3^+$ or the vast majority of it is known or thought to be in diffuse gas; \citep[see][for the argument pertaining to IRAS 08572+3915]{geb06}. In the case of GCS~3-2, however, the sightline passes through both diffuse and dense interstellar gas. \citet{oka05} have determined the column densities in each environment. A further complication is that the lines are Doppler broadened so that the pair is observed as a blend. However, \citet{oka05} also estimated the populations in the (1,1) and (1,0) levels, allowing the equivalent width of the blended pair of lines, which absorb from those levels, to be determined using the formula in \citet{geb96} relating level column density to line equivalent width.  

If the bulk of the H$_3^+$ is in the lowest para and lowest ortho levels, as is generally the case, then the  equivalent width of the line pair represents the total column density of H$_3^+$. One then would naively expect that the equivalent widths in Fig. 2 would scale with one another. NGC~1068 clearly deviates the most from following that trend; the discrepancy is roughly a factor of six. This suggests either that an unusually low fraction of its total  H$_3^+$ column density resides in its two lowest energy levels or that it has a highly enhanced abundance of the carrier of the 3.4-$\mu$m absorption feature, or both. There are several possible explanations for this anomaly, which we now list and briefly discuss.

\begin{enumerate}

\item H$_3^+$  in other energy levels: For cold diffuse gas such as that in front of Cygnus OB2 No. 12 the (1,1) and (1,0) levels  are the only ones that are significantly populated.  In the Galactic center, where the density of the diffuse gas containing the H$_3^+$ is $\sim$100~cm$^{-3}$ and the gas temperature is $\sim$250~K the metastable (3,3) level contains about 23\% of the H$_3^+$  and altogether  all of the levels higher than (1,1) and (1,0) contain about 30\% of the H$_3^+$ \citep{oka05}. Thus the correction, which would move its data point upward by 30\% in Fig. 2  is modest.  In IRAS 08572+3915 the large column densities implied by the strong H$_3^+$ lines suggest  that the H$_3^+$  is distributed along a column length of hundreds of parsecs \citep{geb06}. Much of this  extended region is likely to be at temperatures that are no higher than those in the Galactic center, and thus we believe that the upward correction is also small. 

If the H$_3^+$  observed toward NGC~1068 is in gas physically associated with the continuum source, as we suspect, its temperature could be higher. \citet{gra06} suggest dust temperatures of $\sim$600~K for the dust emitting in the L band. \citet{all97}, observing at $K$, $L$, and $M$, found a mean dust temperature of 750~K for the central $\sim$1-arc second region. \citet{jaf04}, observing near 10~$\mu$m, find a 320~K in the central few parsecs, as do others \citep{mas06, pon08}. Thus a significant portion of the absorbing column may have a higher proportion of H$_3^+$ in higher levels than the Galactic center.  However, due to the wide spacing of rotation energy levels in the ground vibrational state (e.g., the $J$=4 levels have excitation energies $>$1,000~K) as well as depopulation via spontaneous decay of many of the excited rotational states at the low densities of diffuse gas \citep{oka04}, we expect the correction to be less than a factor of two. Thus we do not think that large populations in higher energy levels can explain the discrepant data point in Fig.~2. 

\item Higher gas density and shorter column length: In diffuse gas the density of  H$_3^+$ does not track the total number density of the gas in which it is located, but is a constant whose value depends on the rate of cosmic ray ionization of H$_2$ and the rate of dissociative recombination of H$_3^+$ on electrons \citep{geb99}. Thus the column density of H$_3^+$ scales with column length rather than with gas column density. If the mean density of the diffuse environment in front of the AGN in NGC~1068 is higher than in typical diffuse clouds, the column density of H$_3^+$ there will be lower relative to other species. Likewise, a larger than normal density of free electrons, which is the principal destroyer of H$_3^+$ in diffuse gas, would result in a lower steady state density of H$_3^+$ and also in a lower column density.

\item Low fractional abundance of H$_2$: As discussed in Section 2.3.2 of  \citet{oka13} the abundance of H$_3^+$  is approximately proportional to the square of the fractional abundance of molecular hydrogen, $f$(H$_2$)~=~2$n$(H$_2$)/[$n$(H)~+~2$n$(H$_2$)]. A value of $f$(H$_2$) that is smaller by a factor of 2.5  than in typical Galactic diffuse clouds could explain the low abundance of H$_3^+$.  In NGC~1068 the abundance of H$_2$ might well be lower than normal in the hostile environment close to the AGN.  

\item High abundance of carbon and/or the carrier of the 3.4-$\mu$m feature: 

The 3.4-$\mu$m feature is generally attributed to C-H stretching vibrations in aliphatic hydrocarbons. As such the strength of the feature is sensitive to the carbon abundance and is likely to be greater toward the centers of galaxies, paralleling expected or observed increases in metallicity. In diffuse clouds, where most of the carbon is singly ionized, a higher abundance of carbon also results in a higher electron density and a lower steady state abundance of H$_3^+$.  Judging from Fig. 2, if these effects alone were to account for the discrepancy the carbon abundance in the nucleus of NGC~1068 would need to be several times higher than in the nuclear regions of IRAS 08572+3915 or the Galactic center. We have found no observations of NGC~1068 in the literature that are pertinent to this issue. 

\end{enumerate}

To summarize this subsection, there are several possible explanations for the unusually low value of the ratio of equivalent widths of the H$_3^+$ and 3.4-$\mu$m absorption features on the sightline to the AGN in NGC~1068. One, some, or all of them could contribute and we are unable to select a likely dominant contributor.

\subsection{Outflow and Mass Loss}

The observation of a blue-shifted H$_3^+$ absorption feature at 3.7~$\mu$m indicates that on the line of sight to the nucleus, and thus quite far removed from the axis of the ionization bicone, there exists outflowing gas located close to where $L$-band continuum is formed, at a distance of possibly only a few parsecs or a few tens of parsecs from the central engine. The relationship of this outflow to the apparently much more powerful flow largely within the ionization bicone, which probably originates much closer to the central engine, is unclear. 

Based on the H$_3^+$ absorption feature, a  very crude estimate of the rate of mass outflow can be derived. We based our estimate on (1) the H$_3^+$ being in the same diffuse interstellar environment that produces the 3.4-$\mu$m hydrocarbon feature and (2) our interpretation that  the similar values of  the likely optical depth to the source of the $L$-band continuum and the optical depth estimated from the hydrocarbon feature mean that the diffuse gas is in proximity to the continuum sources rather than at a great distance from them. Using a diffuse cloud gas density  of 100 cm$^{-3}$ and a mean distance of the flow at that density of 20~pc, together with the highly dubious assumption of isotropy of the outflow, we obtain an outflow rate of 1~M$_{\odot}$~yr$^{-1}$.  The rate is proportional to the square of the assumed mean distance, which together with the assumption of spherical symmetry have high uncertainties. Other contributors to the uncertainty  are the outflow velocity, which depends on the choice of systemic velocity, and the mean density of the absorbing diffuse gas. We regard the above value of mass outlaw rate as an order of magnitude estimate only. 

If our estimate of 20 pc for the location of the diffuse gas is roughly correct then the outflow observed in H$_3^+$  exists on distance scales smaller than those probed by the ALMA millimeter and submillimeter spectroscopy of the CND. Based on the analysis of molecular line profiles at distances of 100-200~pc  from the center \citet{gar14} recently deduced the existence of an outflow originating in the AGN with a mass loss rate 1-2 orders of magnitude greater than we derive from the blueshifted H$_3^+$ absorption.  They also measure outflow velocities of up to a few hundred km~s$^{-1}$, considerably higher than seen in H$_3^+$. \cite{bar14} also observe higher velocities, but derive a mass loss rate, based on the kinematics of the ionized gas similar to our rough estimate for the diffuse molecular gas. Unlike the outward flowing H$_3^+$  observed by us, which is far from being along the axis  of the bicone, the outflowing gas that both of these groups have observed is within or at least much closer to the bicone. Much higher velocities, up to 1,000 km~s$^{-1}$,  also are observed from the bicone in numerous optical and infrared emission lines \citep[e.g.,][]{cre00,das06,pon08}.

The differences in mass loss rates derived here and by \citet{gar14} thus may indicate large anisotropies in the outflow, but also could easily be due to time variability both in intensity and in the direction of the outflow.  Approximately 10$^{5}$ years are required for gas moving at 1,000 km~s$^{-1}$ to travel 100~pc (the distance from the central engine to the inner edge of the CND). On the other hand, it seems reasonable that although the outflow is primarily within the bicone, it has an associated and much weaker component in other directions. That portion of the flow would be more shielded from the strong ultraviolet field present in the bicone and would allow gas that on the line of sight to become or to remain partially molecular, although quite close to the AGN. 

\section{Conclusions}

We have detected a weak absorption by blue-shifted H$_3^+$ on the line of sight to the central bright infrared continuum source of the type II Seyfert galaxy NGC~1068.  This is only the second detection of H$_3^+$ in an extragalactic object. Surprisingly H$_3^+$  is the only interstellar molecule in NGC~1068 that has been detected in absorption toward its AGN. Its presence is consistent with the existence in front of the AGN of a significant column density of diffuse gas, previously known to be present by a prominent 3.4-$\mu$m hydrocarbon feature, and also from the absence of absorption in the fundamental band of carbon monoxide.  We tentatively conclude that the diffuse gas is situated within a few tens of parsecs of the central engine and estimate an (isotropic) mass outflow rate of 1~M$_{\odot}$~yr$^{-1}$ with an uncertainty of an order of magnitude. This rate is lower by 1-2 orders of magnitude than the rate recently deduced by \citet{gar14} from ALMA observations of line profiles in the 100-200~pc radius CND, but is roughly consistent with the rate derived by \citet{bar14}. The H$_3^+$ absorption is considerably weaker than expected from the strength of the 3.4-$\mu$m feature. This could be explained by the gas being at a higher temperature, by the fraction of hydrogen in molecular form being 2-3 times lower than typical for diffuse clouds, by an unusually high abundance of carbon in the nucleus, or by the diffuse gas in which the absorption takes place being considerably higher density than standard Galactic diffuse clouds in which H$_3^+$ has previously been detected.

\bigskip\bigskip

{\noindent Acknowledgements}

This paper is based on observations obtained at the Gemini Observatory, which is operated by the Association of Universities for Research in Astronomy, Inc., under a cooperative agreement with the NSF on behalf of the Gemini partnership: the National Science Foundation 
(United States), the National Research Council (Canada), CONICYT (Chile), the Australian Research Council (Australia), Minist\'{e}rio da Ci\^{e}ncia, Tecnologia e Inova\c{c}\~{a}o (Brazil) and Ministerio de Ciencia, Tecnolog\'{i}a e Innovaci\'{o}n Productiva (Argentina).

\end{document}